\documentclass[%
 reprint,
superscriptaddress,
showpacs,
 amsmath,amssymb,
 aps,
floatfix,
]{revtex4-1}

\usepackage{dcolumn}
\usepackage{bm}

\usepackage{graphicx} 
\usepackage{color}

\usepackage{ulem} 

\usepackage[dvipdfmx,
colorlinks=true, citecolor=blue, urlcolor=blue, linkcolor=blue,
setpagesize=false, bookmarks=false
]{hyperref}

\begin{document}
\title{Effect of phase string on single-hole dynamics in the two-leg Hubbard ladder}
\author{Kazuya Shinjo}
\affiliation{Department of Applied Physics, Tokyo University of Science, Tokyo 125-8585, Japan}
\author{Shigetoshi Sota}
\affiliation{Computational Materials Science Research Team,
RIKEN Center for Computational Science (R-CCS), Kobe, Hyogo 650-0047, Japan}
\author{Takami Tohyama}
\affiliation{Department of Applied Physics, Tokyo University of Science, Tokyo 125-8585, Japan}

\date{\today}
             

\begin{abstract} 
Optical measurements in doped Mott insulators have discovered the emergence of spectral weights at mid-infrared (MIR) upon chemical doping and photodoping.
MIR weights may have a relation to string-type excitation of spins, which is induced by a doped hole generating misarranged spins with respect to their sublattice.
There are two types of string effects: one is an $S^z$ string that is repairable by quantum spin flips and the other is a phase string irreparable  by the spin flips.
We investigate the effect of $S^{z}$ and phase strings on MIR weights.
Calculating the optical conductivity of the single-hole Hubbard model in the strong-coupling regime and the $t$-$J$ model on two-leg ladders by using time-dependent Lanczos and density-matrix renormalization group, we find that phase strings make a crucial effect on the emergence of MIR weights as compared with $S^{z}$ strings.
Our findings indicate that a mutual Chern-Simons gauge field acting between spin and charge degrees of freedom, which is the origin of phase strings, is significant for obtaining MIR weights.
Conversely, if we remove this gauge field, no phase is picked up by a doped hole.
As a result, a spin-polaron accompanied by a local spin distortion emerges and a quasiparticle with a cosine-like energy dispersion is formed in single-particle spectral function.
Furthermore, we suggest a Floquet engineering to examine the phase-string effect in cold atoms.
\end{abstract}
\maketitle

%
\section{introduction}\label{sec_1}
After the discovery of the high-temperature superconductivity in cuprate materials, the ground and excited states of hole-doped Mott insulators have been extensively investigated both theoretically and experimentally~\cite{Dagotto1994, Imada1998}.
Nevertheless, we have a long-standing mystery on the fundamental nature of the hole-doped two-dimensional Hubbard model, for example, superconducting correlations~\cite{Qin2020}.
Even if we restrict ourselves to a single-hole problem, there has been a lot of theoretical debates on the nature of the hole surrounded by antiferromagnetic (AFM) spins~\cite{Bulaevskii1968, Shraiman1988, Schmitt-Rink1988, Kane1989, Bonca1989, Martinez1991, Sushkov1999, Brunner2001, Mezzacapo2011}.
There is only few that is exactly known such as the emergence of the Nagaoka ferromagnetic (FM) state~\cite{Nagaoka1966, Kollar1996} in the strong-coupling limit.

Optical measurements have provided important information on the nature of the hole-doped Mott insulators~\cite{Dagotto1994,Phillips2010,Basov2011}.
The most striking feature of charge dynamics emerging in the optical conductivity is dynamical spectral-weight transfer upon chemical doping and photodoping.
Especially, spectral weights at mid-infrared (MIR) are sensitive to dimensionality: the enhancement of the weights at MIR was found in two dimensions, while not in one dimension.
In fact, MIR spectral weights have been observed in two-dimensional high-temperature superconducting material La$_{2-x}$Sr$_{x}$CuO$_{4}$~\cite{Uchida1991} and two-leg ladder material Sr$_{14-x}$Ca$_{x}$Cu$_{24}$O$_{41}$~\cite{Osafune1997}.
Theoretically, MIR weights have been studied in the Hubbard~\cite{Dagotto1992,Tohyama2005} and $t$-$J$~\cite{Stephan1990, Inoue1990, Jaklic2000, Tohyama2004} models in two dimensions.
MIR weights are expected to contain essential information on the dynamical properties of holes in the two-dimensional Mott insulators.

\begin{figure}[t]
  \centering
    \includegraphics[clip, width=20pc]{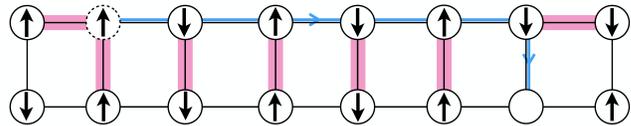}
    \caption{The schematic picture of $S^{z}$ strings. The circles represent lattice points, and up and down arrows show up and down spins, respectively. A $S^{z}$ string is generated when a hole moves along the blue arrows starting from a position denoted by the dotted circle to a position denoted by the empty circle in the N\'{e}el state. As a consequence, the red bonds with spin mismatches increase in proportion to the distance traveled by the hole, resulting in string-type excitations.}
    \label{F0a}
\end{figure}

A hole moving in AFM spin background creates misarranged spins with respect to their sublattice and consequently induce string-type excitation of spins.
MIR weights may have a relation to string structures.
However, it is unclear how MIR weights actually are related to string excitations that are the result of a complex process due to spin-charge coupling.
If a hole moves in an AFM state, it leaves traces of spin mismatches with sub-lattice magnetization as shown in Fig.~\ref{F0a}, which is called $S^{z}$ strings.
If strings consist only of $S^{z}$ strings, which corresponds to the Ising limit, holes are bound to their original position~\cite{Bulaevskii1968,Shraiman1988} by a linear potential proportional to $J_{z}^{2/3}$, where $J_{z}$ is an Ising component of spin-exchange interaction.
The $S^{z}$ strings were later realized~\cite{Schmitt-Rink1988, Kane1989, Bonca1989, Martinez1991} to be relaxed by quantum spin flips once a transverse component of spin-exchange interaction $J_{\perp}$ is introduced.
If spin mismatch generated by the hopping of the hole is repairable by quantum spin flips, a hole becomes a mobile object.
As a result, the mobile hole carries a local spin distortion called spin polaron, which behaves as a quasiparticle with a nonzero spectral weight.
However, this spin-polaron picture is inadequate to describe a hole surrounded by AFM spins, since the hole picks up a nontrivial U(1) phase when it hops in an AFM spin background.
This phase generates another type of strings, that is, $S^{\pm}$ strings caused by transverse spin component, which are regarded as phase strings~\cite{Sheng1996, Weng1997, Weng1999, Weng2011a,Weng2011b}.
It is irreparable by quantum spin flips in contrast to the $S^{z}$ strings. 
It has been further proposed that phase strings, instead of $S^{z}$ strings, are responsible for an intrinsic self-localization of the injected hole in two dimensions~\cite{Zhu2013}.
To investigate string structures in the Mott insulators~\cite{Mazurenko2017, Grusdt2018, Grusdt2019, Fazzini2019, Bohrdt2020, Montorsi2020, Sous2020a, Sous2020b}, highly controlled quantum simulations in cold atoms have recently been proposed.

In this paper, we demonstrate how much $S^{z}$ and phase strings contribute to MIR weights.
We calculate the optical conductivity of the Hubbard model in the strong-coupling regime and $t$-$J$ model by using time-dependent Lanczos and density-matrix renormalization group (DMRG) methods.
We focus on the Mott insulators with a single hole and consider two-leg ladders, which are known to show MIR spectral weights~\cite{Osafune1997, Hashimoto2016}.
Turning on and off the effect of phase strings, we examine how they contribute to MIR weights.
We find that phase strings play an essential role in MIR weights.
MIR weights are crucially suppressed for both the Hubbard and $t$-$J$ models if we remove phase strings.
Although $S^{z}$ strings also contribute to MIR weights, their contribution is smaller than phase strings.
We consider that this is because $S^{z}$ strings can be self-healed via quantum spin flips, while phase strings are not reparable.
Our findings suggest that a mutual Chern-Simons gauge field, which is an elementary force between spin and charge in the phase-string theory~\cite{Weng1997,Kou2005}, is significant for obtaining MIR weights.
This indicates that a hole does not pick up a U(1) phase when moving in AFM spin background if we remove this gauge field.
As a result, we can characterize a doped hole surrounded by AFM spins via a spin-polaron quasiparticle, which has a cosine-like energy dispersion in the single-particle spectral function.

This paper is organized as follows.
We introduce the phase-string theory in Sec.~\ref{sec_2}, which detects the essence of the mutual relationship between holon/doublon and spinon in the doped Mott insulators.
In Sec.~\ref{sec_3}, we introduce the Hamiltonians, where we can switch off phase strings.
Then, we can investigate the phase-string effect in the optical conductivity of the hole-doped Hubbard and $t$-$J$ models on two-leg ladders in Sec.~\ref{sec_4}.
We demonstrate that phase strings play a significant role in MIR weights.
Besides, we confirm that the removal of phase strings induces a spin polaron in the Mott insulators.
We propose the use of Floquet engineering in cold atoms to reduce the effect of phase strings in Sec.~\ref{sec_5}.
Finally, we give a summary of this work in Sec.~\ref{sec_6}.
Note that in this paper, we set the light velocity $c$, the elementary charge $e$, the Dirac constant $\hbar$, and the lattice constant to be 1.

\section{phase-string theory}\label{sec_2}
The slave-boson and slave-fermion mean-field theories are known as the most popular approaches to describe the hole-doped Mott insulators in the strong coupling limit.
In the slave-boson mean-field theory, we treat the spinons as fermions.
There have been many proposals~\cite{Anderson1987, Baskaran1988, Nagaosa1990} of mean-field theories based on the slave boson.
However, there is an inherent problem that this approach does not yield correct AFM correlations at small doping.
This comes from treating spinons as fermions.
Exchanging two same spins gives rise to the sign change of wave function due to the fermionic statistics.
Those redundant and unphysical signs of wave function do not matter if we enforce a strict no-double occupancy constraint.
Once the constraint is relaxed to make a mean-field approximation, the treatment of the signs gives a tricky problem of an overall underestimate of AFM correlation.

From the viewpoint of the Marshall sign rule~\cite{Marshall1955}, which gives an exact description of the ground state of magnetic state at half-filling, a bosonic description of spinons is more natural, where no extra sign is introduced due to the statistics of bosons.
A variational wave function based on the bosonic resonating-valence-bond picture~\cite{Liang1988} based on the slave-fermion mean-field theory~\cite{Arovas1988,Auerbach1988,Auerbach,Kane1989} gives an accurate ground-state energy~\cite{Chen1996}. 
However, the slave-fermion approach fails to describe the ground state away from half-filling.

The fails of these mean-field approaches to the hole-doped Mott insulator imply that doped holes give a singularity that makes the problem beyond description by mean field.
One of the most striking doping effects is the emergence of phase strings~\cite{Sheng1996, Weng1997,Weng1999,Weng2011a,Weng2011b}, which come from spin mismatches due to the hopping of doped holes that cannot be completely repaired through quantum spin flips.

Following Ref.~\cite{Zhang2014}, we consider the slave-fermion representation of the Hubbard model
\begin{align}
\mathcal{H}=\mathcal{H}^{T}+\mathcal{H}^{I}
\end{align}
with
\begin{align}
\mathcal{H}^{T}=&-t_\text{h} \sum_{\langle i,j\rangle, \sigma} \left[ c_{i,\sigma} ^{\dag} c_{j,\sigma} + \text{H.c.}  \right], \\
\mathcal{H}^{I}=& U\sum_{j} \hat n_{j,\uparrow} \hat n_{j,\downarrow},
\end{align}
where $c^\dagger_{i,\sigma}$ is the creation operator of an electron with spin $\sigma (= \uparrow, \downarrow)$ at site $i$ and $n_{i,\sigma}=c^\dagger_{i,\sigma}c_{i,\sigma}$. 
$\langle i,j \rangle$ indicates a nearest-neighbor pair of sites.
$t_\mathrm{h}$ and $U$ are the nearest-neighbor (NN) hopping and on-site Coulomb interaction, respectively.
We take $t_{h}$ to be the unit of energy ($t_{h}=1$).
H.c. is the abbreviation of Hermitian conjugate.
In the standard slave-fermion formalism, $c_{i,\sigma}$ is written as
\begin{align}
c_{i,\sigma}=(-\sigma)^{i} (h_{i}^{\dag}b_{i,\sigma} +\sigma b_{i,-\sigma}^{\dag}d_{i}),
\end{align}
where $h_{i}^{\dag}$ is fermionic holon creation operator, $d_{i}$ is fermionic doublon annihilation operator, and $b_{i,\sigma}$ is spinon annihilation operator treated as (Schwinger) boson.
Note that the staggered phase factor $(-\sigma)^{i}$ is introduced to explicitly take into account the Marshall sign.
There is a constraint on physical Hilbert space as
\begin{align} \label{Eq_constrant}
h_{i}^{\dag} h_{i} + d_{i}^{\dag} d_{i} + b_{i,\uparrow}^{\dag} b_{i,\uparrow} + b_{i,\downarrow}^{\dag} b_{i,\downarrow}=1.
\end{align}
With the slave-fermion representation, we obtain
\begin{align} \label{Eq_ps}
\mathcal{H}^{T}=&-t_\text{h}  \sum_{\langle i,j \rangle} \left[ \mathcal{\hat P}_{i,j} + \mathcal{\hat S}_{i,j}^{\uparrow} -  \mathcal{\hat S}_{i,j}^{\downarrow} \right] + \text{H.c.}, \\
\mathcal{H}^{I}=&U\sum_{j} d_{j}^{\dag}  d_{j},
\end{align}
where 
\begin{align}
\mathcal{\hat P}_{i,j}=&\sum_{\sigma} (b_{i,\sigma}^{\dag} b_{j,-\sigma}^{\dag} h_{i} d_{j} + b_{j,\sigma}^{\dag} b_{i,-\sigma}^{\dag} h_{j} d_{i})
\end{align}
creates a spinon pair and annihilates holon and doublon pair and 
\begin{align}
\mathcal{\hat S}_{i,j}^{\sigma}=&b_{i,\sigma}^{\dag} b_{j,\sigma} h_{j}^{\dag} h_{i} + b_{i,\sigma}^{\dag} b_{j,\sigma} d_{j}^{\dag} d_{i}
\end{align}
swaps a holon and doublon with a spinon with spin $\sigma$. 

We span the Hilbert space with the basis $\{ |\alpha \rangle = d_{l_{1}}^{\dag}\cdots h_{m_{1}}^{\dag}\cdots b_{i_{1},\uparrow}^{\dag}\cdots b_{j_{1},\downarrow}^{\dag}\cdots |0\rangle \}$.
Here, the constraint Eq.~(\ref{Eq_constrant}) is always satisfied.
Making a high-temperature expansion to all orders and inserting completeness $\sum_{\alpha}|\alpha \rangle \langle \alpha |=1$,
the partition function is given as
\begin{align}
Z=\text{Tr} e^{-\beta \mathcal{H}}=\sum_{n=0}^{\infty} \frac{\beta^{n}}{n!} \sum_{\alpha_{i}} \prod _{i=0}^{n-1} \langle \alpha_{i+1}|(-\mathcal{H})|\alpha_{i}\rangle
\end{align}
with $|\alpha _{n}\rangle=|\alpha _{0}\rangle$.
Here, $\beta$ is inverse temperature.
Then, the partition function is represented as a summation of a closed paths $c$'s as
\begin{align}
Z=\sum_{c} (-1)^{N[c]} W_{h}[c],
\end{align}
where $W_{h}[c]$ is a positive-definite weight (see Appendix of Ref.~\cite{Zhang2014} for detail).
The sign factor is given by
\begin{align}
N[c]=N_{h}^{h}[c] + N_{d}^{d}[c] + N_{h}^{\downarrow}[c] + N_{d}^{\downarrow} [c].
\end{align}
$N_{h(d)}^{h(d)}[c]$ denotes the total number of the holon-holon (doublon-doublon) exchange on a path $c$ and $N_{h(d)}^{\downarrow}[c]$ the total number of the holon-$\downarrow$ spin exchange on a path $c$. 
If there is only one holon (doublon) in the Mott insulator, a holon (doublon) does not exchange with another holon (doublon), giving rise to $N_{h(d)}^{h(d)}[c]=0$, i.e. $N[c]=N_{h}^{\downarrow}[c] + N_{d}^{\downarrow} [c]$.
Note that the number of exchanges between holon and doublon are always even on a path $c$ and $N_{h(d)}^{d(h)}[c]$ does not contribute to $N[c]$.
In the limit of $U/t_\text{h}\gg 1$, doubly-occupied sites vanish giving rise to $N[c]=N_{h}^{\downarrow}[c]$, which is realized in the one-hole doped $t$-$J$ model.

\begin{figure}[t]
  \centering
    \includegraphics[clip, width=20pc]{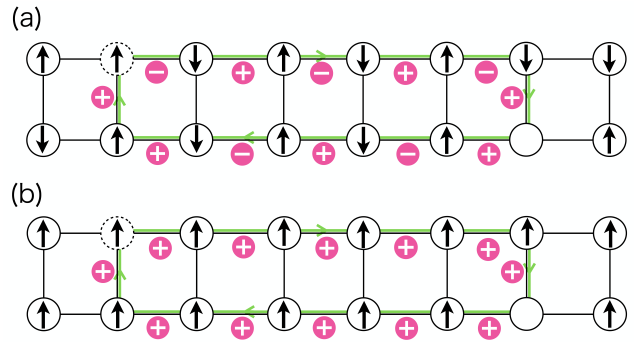}
    \caption{The schematic picture of phase strings. The circles represent lattice points, and up and down arrows show up and down spins, respectively. The sequence of signs ``$+$'' and ``$-$'' is obtained along the green closed path when a hole moves from the dotted circle to the solid empty circle in spin backgrounds. A phase string is given by multiplying all the signs along the green path. (a) The background spins form the N\'{e}el state. (b) The background spins form FM state.}
    \label{F0b}
\end{figure}

Figures~\ref{F0b}(a) and \ref{F0b}(b) illustrate phase strings acquired by a hole moving along a closed loop. The background spins form the N\'{e}el state and FM state in Figs.~\ref{F0b}(a) and \ref{F0b}(b), respectively. The phase interference due to phase strings is always constructive in Fig.~\ref{F0b}(b). However, that is nontrivial in Fig.~\ref{F0b}(a) since a local quantum spin flip can change a global sign $(-1)^{N[c]}$.
Since the probability for the spin configuration in Fig.~\ref{F0b}(b) in the partition function $Z$ is extremely small for finite spin-exchange interaction, frustration from phase strings plays an essential role in characterizing a mutual relationship between charge and spin degrees of freedom.
If we remove phase strings, negative signs in Fig.~\ref{F0b}(a) change to positive signs. Then, we obtain a state with a phase-string structure equivalent to that formed by the Nagaoka polaron.

Phase string $(-1)^{N[c]}$ is regarded as a generalized Berry's phase for a state $|\alpha \rangle$ adiabatically moving on a given path $c$.
We should embed this fluctuating phase in the wave function ansatz as \textit{a priori} in constructing variational and mean-field states.
The phase-string theory gives this type of treatment of holon/doublon and spinon.
Then, the elementary force between holon/doublon and spinon is mediated by a mutual Chern-Simons gauge fields, while it is mediated by U(1)~\cite{Lee1992} or SU(2)\cite{Wen1996, Lee1998} gauge field in the slave-boson approach.
Controlling the strength of the gauge field, dynamical quantities such as optical conductivity\cite{Gu2007} and dynamical spin susceptibility\cite{Chen2005,Gu2005} of the $t$-$J$ model have been examined in the phase-string theory.
Within the mean-field approximation, a holon propagator evaluated by calculating the average of the phase string on all possible paths $c$'s shows an exponential decay in space giving rise to localization phenomenon~\cite{Weng2001}.
It is fundamentally different from the power-law decay in space expected when the hole behaves like a well-defined quasiparticle.

Combining the phase-string theory with a nonperturbative numerical technique such as DMRG gives essential information on the mutual relationship between holon/doublon and spinon in the hole-doped Mott insulators.
For example, the self-localization of a hole in the Mott insulators is a longstanding controversial problem.
The study combining the phase-string theory with DMRG has suggested the emergence of self-localization in the even- and odd-leg $t$-$J$ ladders~\cite{Zhu2013}.
Besides, the effect of phase strings on superconducting pair correlations has been investigated in the $t$-$J$ model~\cite{Jiang2019}.
Transforming the Hamiltonian to turn on and off the phase strings, we can examine the role of string structures in the Mott insulators with nonperturbative numerical methods.

\section{model Hamiltonians}\label{sec_3}
If we change the sign of hopping involving the exchange of a holon/doublon and a $\downarrow$ spin accompanied by $\mathcal{\hat S}_{i,j}^{\downarrow}$ in Eq.~(\ref{Eq_ps}), we can cancel out phase strings. After applying such manipulation, the Hamiltonian of the phase-string-removed Hubbard model is written as~\cite{Zhu2016,He2016}
\begin{align}
\mathcal{H}_\text{r}=\mathcal{H}^{T}_\text{r}+\mathcal{H}^{I}
\end{align}
with
\begin{align}
\mathcal{H}^{T}_\text{r}=&-t_\text{h} \sum_{\langle i,j\rangle, \sigma} \left[ c_{i,\sigma} ^{\dag} c_{j,\sigma} (\sigma \hat P + \hat Q) + \text{H.c.}  \right],
\end{align}
where $\hat P$ is a projection operator onto hopping processes involving the exchanges between a $\sigma$ spin with a holon or doublon, $\hat Q$ is a projection operator onto hopping processes involving exchange between singly occupied sites and exchange between a holon and a doublon, and
$\sigma=1$ ($-1$) is a spin-dependent sign corresponding to $\uparrow$ ($\downarrow$) spin.

The removal of phase strings is also done by introducing the transformation of an operator called U(1) nonlinear (NL) transformation as
\begin{align}
c_{j,\uparrow}  \xrightarrow{\text{U(1)}_\text{NL}} & c_{j,\uparrow}, \\ 
c_{j,\downarrow}  \xrightarrow{\text{U(1)}_\text{NL}} & (-1)^{ j} \left[ c_{j,\downarrow} (1-\hat n_{j,\uparrow}) - c_{j,\downarrow} \hat n_{j,\uparrow} \right] \nonumber \\
=&(-1)^{ j}(-1)^{ \hat n_{j,\uparrow}} c_{j,\downarrow}.
\end{align}
Since we obtain
\begin{align}
c_{i,\uparrow}^{\dag} c_{j,\uparrow} \xrightarrow{\text{U(1)}_\text{NL}} & c_{i,\uparrow}^{\dag} c_{j,\uparrow}, \\
c_{i,\downarrow}^{\dag} c_{j,\downarrow} \xrightarrow{\text{U(1)}_\text{NL}} &
-(1-\hat n_{i,\uparrow})c_{i,\downarrow}^{\dag} c_{j,\downarrow} (1-\hat n_{j,\uparrow}) \nonumber\\
&-n_{i,\uparrow}c_{i,\downarrow}^{\dag} c_{j,\downarrow}\hat n_{j,\uparrow} \nonumber \\
&+\hat n_{i,\uparrow}c_{i,\downarrow}^{\dag} c_{j,\downarrow}(1-\hat n_{j,\uparrow})\nonumber \\
&+(1-\hat n_{i,\uparrow}) c_{i,\downarrow}^{\dag} c_{j,\downarrow} \hat n_{j,\uparrow}\nonumber \\
=& -(-1)^{ (\hat n_{i,\uparrow}+\hat n_{j,\uparrow})} c_{i,\downarrow}^{\dag} c_{j,\downarrow} ,
\end{align}
and 
\begin{align}
\hat n_{i,\sigma}  \xrightarrow{\text{U(1)}_\text{NL}}  \hat n_{i,\sigma}
\end{align}
for the bipartite lattice, the phase-string-removed Hamiltonian $\mathcal{H}$ is rewritten by introducing
\begin{align}\label{Eq_SS}
\mathcal{H}_\text{r}
=& -t_\text{h} \sum_{\langle i,j\rangle, \sigma}  \left[ e^{i\mathcal{A}_{i,j}^{\sigma}}  c_{i,\sigma}^{\dag} c_{j,\sigma} + \text{H.c.}  \right] 
+ U\sum_{j} \hat n_{j,\uparrow} \hat n_{j,\downarrow}
\end{align}
with 
\begin{align}
\mathcal{A}_{i,j}^{\sigma}= \phi^{\sigma}(\hat n_{i,-\sigma}+\hat n_{j,-\sigma}+1),
\end{align}
where we take $(\phi^{\uparrow},\phi^{\downarrow})=(0,\phi)$.
Unless otherwise noted, we use $\phi=\pi$, by which the phase-string effect is completely removed.
$\mathcal{A}_{i,j}^{\sigma}$ is interpreted as the gauge field that cancels out the Aharonov-Bohm phases arising from a flux bounded to down spins.
Note that the U(1) group of $\text{U(1)}_\text{NL}$ transformation does not have local gauge symmetry for $\mathcal{H}$ when $t_\text{h}\neq0$~\cite{Ostlund1991}.

Using the Schrieffer-Wolf transformation, we obtain the $t$-$J$ model as an effective model in the strong coupling limit denoted as
\begin{align}
\mathcal{H}^{t\text{-}J} =& -t_\text{h} \sum_{\langle i,j\rangle,\sigma} \left[\bar c_{i,\sigma} ^{\dag} \bar c_{j,\sigma} + \text{H.c.}\right] 
+ J\sum_{\langle i,j \rangle} \bm{S}_{i} \cdot \bm{S}_{j}, \\
\mathcal{H}^{t\text{-}J}_\text{r}=& -t_\text{h} \sum_{\langle i,j\rangle,\sigma} \sigma \left[ \bar c_{i,\sigma} ^{\dag} \bar c_{j,\sigma} + \text{H.c.}\right] 
+ J\sum_{\langle i,j \rangle} \bm{S}_{i} \cdot \bm{S}_{j} 
\end{align}
with $\bar c_{j,\sigma} = c_{j,\sigma} (1-\hat n_{j,-\sigma})$ and $J=4t_{h}^{2}/U$.
The $\text{U(1)}_\text{NL}$ transformation is a generalization of the transformation $\bar c_{j,\downarrow} \rightarrow e^{i\pi j} \bar c_{j,\downarrow}$ given for the $t$-$J$ model~\cite{Jiang2019} to remove the phase-string effect.
For the Hubbard model, charge fluctuations in the upper Hubbard band should be additionally considered to remove phase strings.

We note here that the Hamiltonian (\ref{Eq_SS}) is regarded as that of a generalized Schulz-Shastry model $\mathcal{H}_\text{gSS}$ if $\phi^{\sigma}$ is taken as a free parameter~\cite{Amico1998,Osterloh2000}.
If we take $(\phi^{\uparrow},\phi^{\downarrow})=(0,\phi)$ [$(\phi^{\uparrow},\phi^{\downarrow})=(\phi,-\phi)$], $\mathcal{H}_\text{gSS}$ is reduced to $\mathcal{H}_\text{r}$ ($\mathcal{H}_\text{SS}$), where $\mathcal{H}_\text{SS}$ is the Hamiltonian of the Schulz-Shastry model~\cite{Schulz1999,Kundu1998}.
The Schulz-Shastry model, where particles of one spin orientation give rise to an effective Aharonov-Bohm flux acting on the other species, is also known as a specific case of the two-component anyon-Hubbard model~\cite{Cardarelli2016}.
$\mathcal{H}_\text{SS}$ in one-dimensional chain is known as an integrable model.
Absorbing the spin-dependent correlated hopping into a twisted boundary condition~\cite{Shastry1990} with a unitary transformation $\mathcal{U}=\mathcal{U}_{2}\mathcal{U}_{1}$, where 
$
\mathcal{U}_{1}=\exp \left( i\sum_{l>m} \phi [\hat n_{l,\uparrow}\hat n_{m,\downarrow} - \hat n_{m,\uparrow}\hat n_{l,\downarrow}]  \right)
$
and
$
\mathcal{U}_{2}=\prod_{l=0}^{L-1} \exp \left( \frac{2i\phi l (\hat N_{\uparrow}\hat n_{l,\downarrow}-\hat N_{\downarrow}\hat n_{l,\uparrow})}{L} \right)
$
with $\hat N_{\sigma}=\sum_{j} \hat n_{j,\sigma}$,
the problem is reduced to solving the Schr\"{o}dinger equation of the usual Hubbard model.
However, in the case of $\mathcal{H}_\text{r}$, we cannot obtain a solvable Schr\"{o}dinger equation by the Bethe ansatz~\cite{Osterloh2000}.

$\mathcal{H}_\text{r}^{T}$ is also represented by explicitly introducing  correlated hopping as
\begin{align}\label{Eq_corrhop}
\mathcal{H}^{T}_\text{r}=&- \sum_{\langle i,j \rangle, \sigma} \Bigl\{ t^{\sigma} c_{i,\sigma} ^{\dag} c_{j,\sigma}\nonumber \\
&- \Delta t^{\sigma} \left[ \hat n_{i,-\sigma}c_{i,\sigma} ^{\dag} c_{j,\sigma} + c_{i,\sigma} ^{\dag} c_{j,\sigma} \hat n_{j,-\sigma}  \right]\nonumber \\
&+ 2 t^{\sigma}_\text{ex} \hat n_{i,-\sigma} c_{i,\sigma} ^{\dag} c_{j,\sigma} \hat n_{j,-\sigma} 
+\text{H.c.}
 \Bigr\}
\end{align}
with $t^{\sigma}=\sigma t_\text{h}$, $\Delta t^{\sigma}=(\sigma-1)t_\text{h}$, and $t^{\sigma}_\text{ex}=\Delta t^{\sigma}$.
When the correlated hopping is introduced, the Hartree-Fock approximation is given as~\cite{Foglio1979, Amadon1996, Arrachea1997,Aligia1998,Mizia2007,Gorski2011}
\begin{align}
&\hat n_{i,-\sigma} c_{i,\sigma} ^{\dag} c_{j,\sigma} + c_{i,\sigma} ^{\dag} c_{j,\sigma} \hat n_{j,-\sigma} \nonumber \\
& \simeq (n_{i,-\sigma} +n_{j,-\sigma})c_{i,\sigma} ^{\dag} c_{j,\sigma}
+ (\hat n_{i,-\sigma} + \hat n_{j,-\sigma}) I_{\sigma},
\end{align}
\begin{align}
&\hat n_{i,-\sigma} c_{i,\sigma} ^{\dag} c_{j,\sigma} \hat n_{j,-\sigma} \nonumber \\
 \simeq& 
\left(n_{i,-\sigma} n_{j,-\sigma} - I_{-\sigma}^{2} \right) c_{i,\sigma} ^{\dag} c_{j,\sigma}
 -2I_{\sigma}I_{-\sigma} c_{i,-\sigma} ^{\dag} c_{j,-\sigma},
\end{align}
where $n_{j,\sigma}=\langle \hat n_{j,\sigma} \rangle$ and $I_{\sigma}=\langle c_{\sigma,i} ^{\dag} c_{\sigma,j} \rangle$.
Then, we obtain
\begin{align}\label{Eq_HF}
\mathcal{H}^{T}_\text{r}\simeq&- \sum_{\langle i,j \rangle, \sigma}  t^{\sigma}_\text{eff} (c_{i,\sigma} ^{\dag} c_{j,\sigma} + \text{H.c.}) + \sum_{i,\sigma} M_{i}^{\sigma} \hat n_{i,\sigma},
\end{align}
where the effective hopping $t^{\sigma}_\text{eff}=t^{\sigma}b^{\sigma}$ with the bandwidth factor $b^{\sigma}$ given by
\begin{align}
b^{\sigma} =& 1-\frac{\Delta t^{\sigma}}{t^{\sigma}} (n_{i,-\sigma}+n_{j,-\sigma}) \nonumber \\
&+ \frac{2t^{\sigma}_\text{ex}}{t^{\sigma}} (n_{i,-\sigma}n_{j,-\sigma}-I_{-\sigma}^{2} -2I_{\sigma} I_{-\sigma}).
\end{align}
The effective molecular field $M_{i}^{\sigma}$ is given as
\begin{align}
M_{i}^{\sigma} = 2z\Delta t^{\sigma} I_{-\sigma} -4t^{\sigma}_\text{ex} I_{-\sigma} \sum_{j\in \text{NN}(i)} n_{j,\sigma}
\end{align}
for the NN sites $i$ and $j$.
According to the approximated Hamiltonian (\ref{Eq_HF}), we can remove phase strings by introducing (i) a spin-dependent bandwidth factor by $b^{\sigma}$ and (ii) a potential $M_{i}^{\sigma}$ acting only on down spin, which cancels out the sign structure hidden in the Hubbard model.

\begin{figure}[t]
  \centering
    \includegraphics[clip, width=20pc]{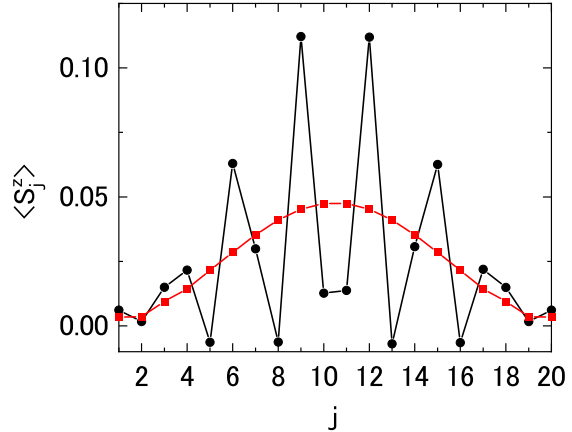}
    \caption{Real space distribution of spin $\langle S_{i}^{z} \rangle$ along the chain of the single-hole two-leg Hubbard ladder with $L_x=20$ for $U=10$. Black points are data for the Hubbard model. Red points are for the Hubbard model without phase-string effects.}
    \label{F1}
\end{figure}

More generally, correlated hoppings are introduced if we consider a strong electron-boson coupling in the antiadiabatic limit of an infinite boson frequency ~\cite{Mahan, Hirsch2001, Hirsch2002} via the Lang-Firsov transformation~\cite{Lang1962}.
In such a case, a polaron quasiparticle picture gives a good description of the system~\cite{Mahan}.
Based on a spin-polaron picture, a bubble of polarized spins called the Nagaoka polaron forms around the hole while the staggered magnetic order realizes further away from the hole~\cite{Nagaoka1966}. 
Although we expect this picture to hold for large $U$~\cite{White2001}, the removal of phase strings via spin-dependent correlated hopping as shown in Eq.~(\ref{Eq_corrhop}) makes a spin-polaron picture valid even when $U$ is an intermediate value.
If a spin-polaron quasiparticle is well defined, we expect a reduction of the bandwidth~\cite{Kollar2001}, which is directly incorporated with the bandwidth factor $b^{\sigma}$.
The band narrowing lowers the amount of energy that is necessary for a FM spin polarization.
We show in Fig.~\ref{F1} the spatial distribution of $\langle S_{j}^{z} \rangle$ for the one-hole-doped two-leg Hubbard ladder with $(L_{x},L_{y})=(20,2)$.
Here, we define the $x$ and $y$ directions as along the leg and rung of the ladder, respectively.
We set the number of sites $L=L_{x}L_{y}$ with $L_{x}$ sites along the leg and $L_{y}$ sites along the rung.
The black points are for the Hubbard model with $U=10$, while the red ones are the same as the black ones but phase strings are removed.
The calculation is performed by DMRG keeping 2000 density-matrix eigenstates, giving rise to the truncation error less than $10^{-9}$.
We find that the staggered spin modulation disappears with the removal of phase strings, which indicates the formation of the Nagaoka polaron even for the intermediate $U$.
The spin modulation has been reported in Refs.~\cite{Zhu2015, Zhu2016}.

\section{Results and discussions}\label{sec_4}

\subsection{Optical conductivity}\label{sec_4a}
We demonstrate and discuss how the $S^{z}$ and phase strings contribute to MIR weights by calculating the optical conductivity of the Hubbard and $t$-$J$ models with single hole on the two-leg ladder.
Since the optical conductivity is a linear response of an electric current to an external spatially homogeneous electric field, we calculate the time-evolution of electric current $\bm{j}^{c}(t)\equiv \langle \frac{\partial H}{\partial \bm{\mathcal{A}}(t)} \rangle$ after applying a gauge field whose vector potential is written as $\bm{\mathcal{A}}(t)$.
The gauge field applied along the chain ($x$ direction) can be incorporated via the Peierls substitution in the hopping terms as $c_{i,\sigma}^\dag c_{j,\sigma} \rightarrow e^{i\bm{\mathcal{A}}(t)\cdot \bm{R}_{ij}}c_{i,\sigma}^\dag c_{j,\sigma}$ with $\bm{\mathcal{A}}(t)=\left(\mathcal{A}_x(t),0\right)$ and $\mathcal{A}_x(t)=\mathcal{A}_0 e^{-(t-t_0)^2/(2t_d^2)} \cos [\Omega (t-t_0)]$.
Here, we set $\bm{R}_{ij} = \bm{R}_{i} -\bm{R}_{j}$.
We obtain the optical conductivity $\sigma(\omega) = j_{x}^{c}(\omega) / \left[i(\omega +i\eta)L\mathcal{A}_{x}(\omega)\right]$, where $\mathcal{A}_{x}(\omega)$ and $j_{x}^{c}(\omega)$ are the Fourier transforms of $\mathcal{A}_{x}(t)$ and the current along the $x$ direction, respectively.
Here, the parameters of the gauge field are $\mathcal{A}_0=0.001$, $t_d=0.02$, $\Omega=10$, and $t_0=1$.
The time-dependent wave function is obtained by the time-dependent Lanczos for $L_{x}\leq 6$ (see Appendix~\ref{Aa}).
For larger systems, we use the time-dependent DMRG (see Appendix~\ref{Ab}).
We employ open boundary conditions and keep 50 Lanczos bases for the Lanczos method and 2000 density-matrix eigenstates for DMRG method.
Since we focus on the linear response regime by taking small $\mathcal{A}_{0}$, we can obtain time-dependent wave functions using DMRG with high accuracy comparable with obtaining ground-state wave functions.

\begin{figure}[t]
  \centering
    \includegraphics[clip, width=20pc]{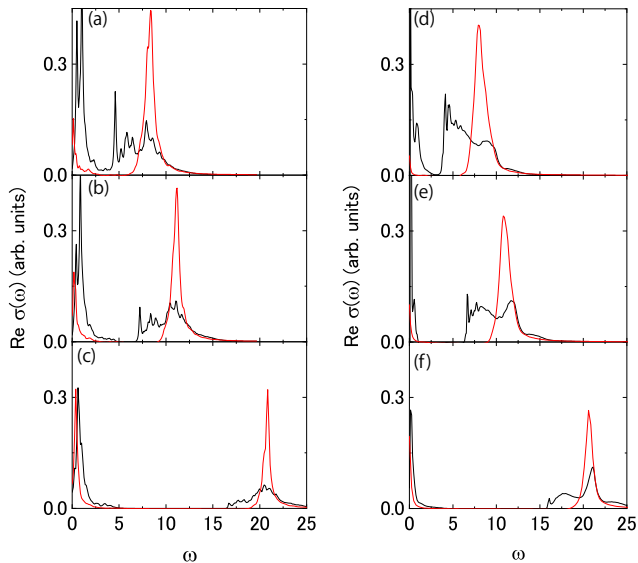}
    \caption{$\text{Re} \sigma (\omega)$ of the single-hole Hubbard model for [(a) and (d)] $U=7$, [(b) and (e)] $U=10$, and [(c) and (f)] $U=20$. (a)-(c) Time-dependent Lanczos method with $L_{x}=6$. (d)-(f) Time-dependent DMRG method with $L_{x}=20$. Black and red lines are the results with and without the phase-string effect, respectively.}
    \label{F2}
\end{figure}

We show $\text{Re}\sigma (\omega)$ of the Hubbard model with single hole for $L_{x}=6$ in Figs.~\ref{F2}(a), \ref{F2}(b), and \ref{F2}(c) for $U=7$, 10, 20, respectively as black lines.
At half-filling, the spectral weights only exist above the Mott gap at $\omega=4.46$, $7.04$, and $16.5$ for $U=7$, 10, and 20, respectively.
Upon doping, the spectral weights above the Mott gap decrease.
The decreased weights are distributed on the small $\omega$, which can be divided into the Drude and MIR weights.
We see that the Drude component is not at $\omega=0$ but finite $\omega$, which is due to the finite size effect in the open boundary condition.
The red lines in Figs.~\ref{F2} (a)-(c) represent data where the effect of phase strings is removed.
We find that the structure above the Mott gap is narrower when phase strings are removed.
This narrowing also occurs in the case of the half filling, which is consistent with the single-particle spectral function discussed in Sec.~\ref{sec_4b}.

To avoid the finite size effect, we calculate $\text{Re}\sigma (\omega)$ for $L_{x}=20$ using time-dependent DMRG as shown in Figs~\ref{F2}(d), \ref{F2}(e), and \ref{F2}(f) for $U=7$, 10, and 20, respectively.
We have performed calculations for other systems with $L_{x}<20$, and have confirmed that the following arguments also hold implying small finite-size effects.
The meaning of the black and red lines is the same as in Figs.~\ref{F2} (a)-(c).
With changing $L_{x}$ from 6 to 20, the peak positions of the spectral weights at low $\omega$ containing the Drude and MIR components shift to the lower energy side.
We find that MIR peaks at $\omega=0.9$ and 0.6 for $U=7$ and 10, respectively.
Since we expect the peak positions of MIR peaks to be $\omega \propto J$, MIR peak for $U=20$ may be located at small $\omega$ and is difficult to be distinguished from the Drude peak.
We find that the removal of phase strings reduces the spectral weights at MIR.
The difference of spectral weights at MIR between black and red lines is apparent for intermediate $U$, while that will be small for large $U$.
This is reasonable since phase strings make no effect on $\text{Re}\sigma (\omega)$ for $U\rightarrow \infty$, i.e., $J\rightarrow 0$.

\begin{figure}[t]
  \centering
    \includegraphics[clip, width=20pc]{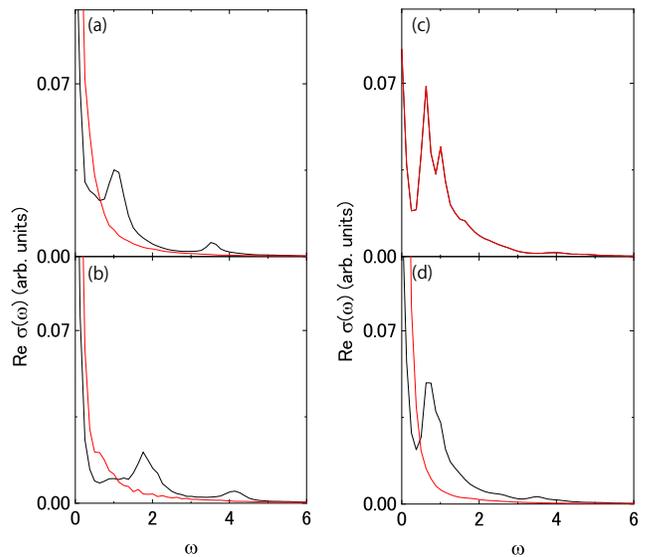}
    \caption{$\text{Re} \sigma (\omega)$ of the single-hole [(a) and (b)] $t$-$J$, (c) $t$-$J_{z}$, and (d) $t$-$J_{\perp}$ models calculated by the time-dependent DMRG with $L_{x}=20$ for (a) $J=0.5$, (b) $J=1$, (c) $J_{z}=0.5$, and (d) $J_{\perp}=0.5$. Black and red lines are the results with and without the phase-string effect, respectively.}
    \label{F3}
\end{figure}

We further examine $\text{Re}\sigma (\omega)$ of the $t$-$J$ model with $L_{x}=20$ shown in Figs.~\ref{F3}(a) and \ref{F3}(b) for $J=0.5$ and 1, respectively.
We find that MIR peaks at $\omega=1.1$ and 1.8 for $J=0.5$ and 1, respectively, as expected from $\omega_\text{MIR} \propto 1/J$. 
Similar to $\text{Re} \sigma (\omega)$ of the Hubbard model, the removal of phase strings suppresses MIR peaks as shown in red lines of Figs.~\ref{F3}(a) and \ref{F3}(b).
Even after removing phase strings, we find small spectral weights at MIR.
We consider that these weights come from the $S^{z}$ strings.

In order to distinguish the effects of $S^{z}$ and phase strings on $\text{Re} \sigma (\omega)$, we consider the $t$-$J_{z}$ and $t$-$J_{\perp}$ models defined as
\begin{align}
\mathcal{H}^{t\text{-}J_z} =& -t_\text{h} \sum_{\langle i,j\rangle, \sigma} \left[\bar c_{i,\sigma} ^{\dag} \bar c_{j,\sigma} + \text{H.c.}\right] + J_{z}\sum_{\langle i,j \rangle} S_{i}^{z}S_{j}^{z}, \\
\mathcal{H}^{t\text{-}J_{\perp}} =& -t_\text{h} \sum_{\langle i,j\rangle,\sigma} \left[ \bar c_{i,\sigma} ^{\dag} \bar c_{j,\sigma} + \text{H.c.}\right] \nonumber \\
&+ J_{\perp}\sum_{\langle i,j \rangle}  \left[S_{i}^{x}S_{j}^{x}+S_{i}^{y}S_{j}^{y}\right].
\end{align}
Phase-string-removed Hamiltonians for $\mathcal{H}^{t\text{-}J_{z}}$ and $\mathcal{H}^{t\text{-}J_{\perp}}$ are represented as $\mathcal{H}^{t\text{-}J_{z}}_\text{r}$ and $\mathcal{H}^{t\text{-}J_{\perp}}_\text{r}$, respectively.
Black (red) lines in Figs. \ref{F3}(c) and \ref{F3}(d) indicate $\text{Re} \sigma (\omega)$ for $\mathcal{H}^{t\text{-}J_{z}}$ ($\mathcal{H}^{t\text{-}J_{z}}_\text{r}$) and $\mathcal{H}^{t\text{-}J_{\perp}}$ ($\mathcal{H}^{t\text{-}J_{\perp}}_\text{r}$), respectively.
$\text{Re} \sigma (\omega)$ of $\mathcal{H}^{t\text{-}J_{z}}_\text{r}$ and $\mathcal{H}^{t\text{-}J_{z}}$ in Fig.~\ref{F3}(c) are exactly the same (the black line overlaps with the red one) and have MIR peaks at $\omega=0.75$ and 1.25.
These peak positions are consistent with $\omega=\frac{3}{2}J_{z}$ and $\omega=\frac{5}{2}J_{z}$, respectively~\cite{Poilblanc1993}.
The agreement between $\text{Re} \sigma (\omega)$ of $\mathcal{H}^{t\text{-}J_{z}}$ and that of $\mathcal{H}^{t\text{-}J_{z}}_\text{r}$ is easily understood because the $\text{U(1)}_\text{NL}$ transformation for the $t$-$J$ model operated to remove phase strings does not change $S_{i}^{z}S_{j}^{z}$.
In contrast, for the $t$-$J_{\perp}$ model where the strings consist only of phase strings, we can completely eliminate the effect of strings by removing phase strings via the $\text{U(1)}_\text{NL}$ transformation.
We show that MIR peak at $\omega=0.7$ indicated in the black line of Fig.~\ref{F3}(d) disappears after removing phase strings as shown in the red line.
Since there is no string effect in $\mathcal{H}^{t\text{-}J_{\perp}}_\text{r}$, $\text{Re} \sigma (\omega)$ at low $\omega$ comes only from the Drude peak with damping.
The charge and spin degrees of freedom in the ground state of $\mathcal{H}^{t\text{-}J_{\perp}}_\text{r}$  seem to be completely separated as in the one-dimensional $t$-$J$ model.

We next discuss how much $S^{z}$ strings contribute to spectral weights at MIR.
The difference between $\text{Re} \sigma (\omega)$ of $\mathcal{H}^{t\text{-}J}_\text{r}$ [the red line in Fig.~\ref{F3}(a)] and that of $\mathcal{H}^{t\text{-}J_{\perp}}_\text{r}$ [the red line in Fig.~\ref{F3}(d)] indicates a contribution from $S^{z}$ strings to MIR weights for the $t$-$J$ model with $J=0.5$.
Comparing it with a contribution from phase strings indicated by the difference between $\text{Re} \sigma (\omega)$ of $\mathcal{H}^{t\text{-}J}$ [the black line in Fig.~\ref{F3}(a)] and that of $\mathcal{H}^{t\text{-}J}_\text{r}$ [the red line in Fig.~\ref{F3}(a)], we find that contribution from $S^{z}$ strings is smaller than that from phase strings.
We consider that this is due to the fact that the $S^{z}$ strings can be self-healed via quantum spin flips, while the phase strings are not.
Therefore, we consider that a mutual Chern-Simons gauge field acting between spin and charge degrees of freedom, which makes phase strings irreparable, is crucial for explaining the origin of MIR weights in the doped Mott insulators.

%
%
\subsection{Single-particle spectral function}\label{sec_4b}
\begin{figure*}[tb]
  \centering
    \includegraphics[clip, width=42pc]{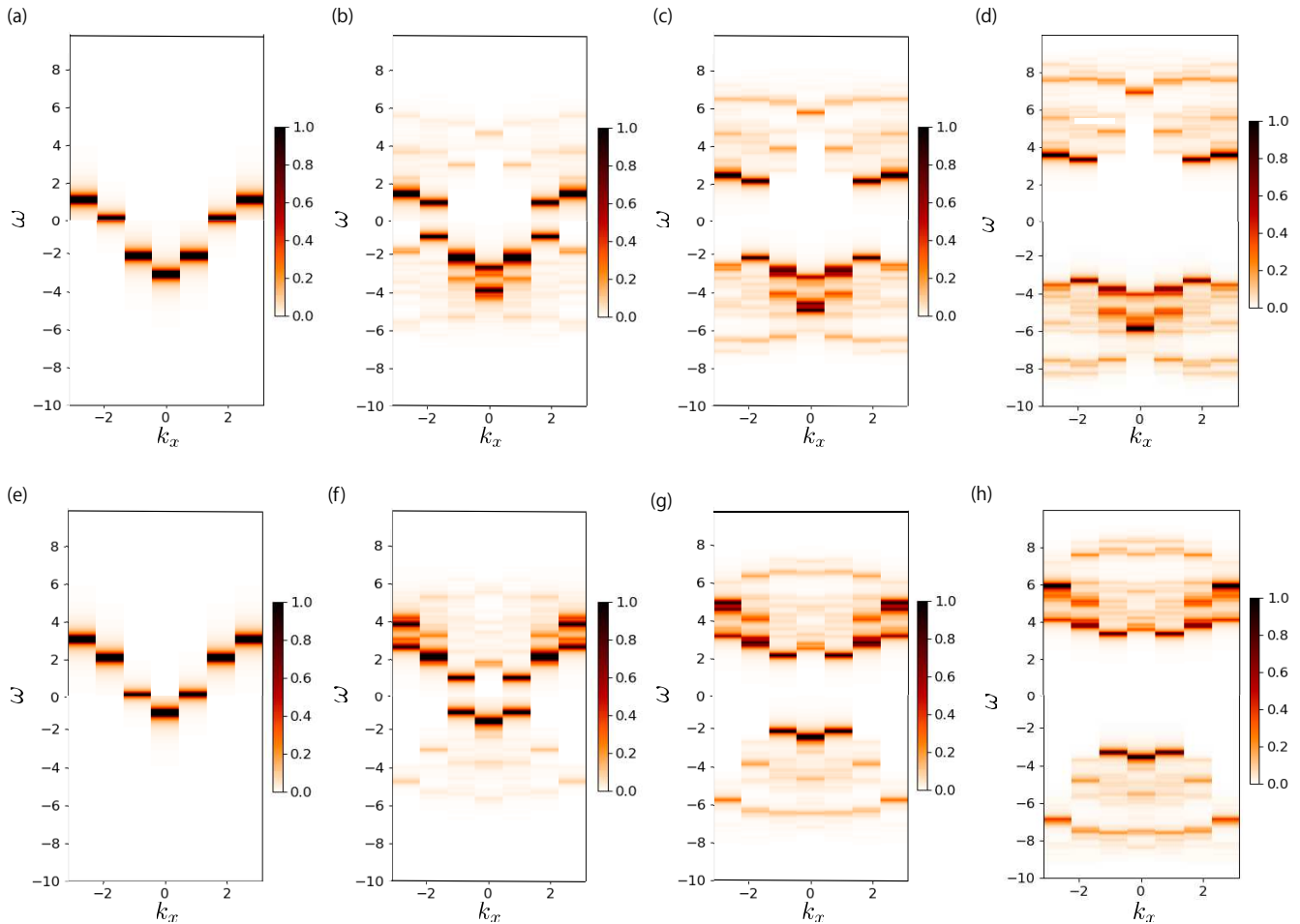}
    \caption{$A_{0}(k_{x},\omega)$ and $A_{\pi}(k_{x},\omega)$ of the Hubbard model at half filling with [(a) and (e)] $U=0$, [(b) and (f)] $U=4$, [(c) and (g)] $U=7$, and [(d) and (h)] $U=10$ calculated by the time-dependent Lanczos with $L_{x}=6$. Upper [(a)-(d)] and lower [(e)-(h)] figures are for $k_{y}=0$ and $\pi$ sectors, respectively.}
    \label{F4}
\end{figure*}

\begin{figure*}[tb]
  \centering
    \includegraphics[clip, width=42pc]{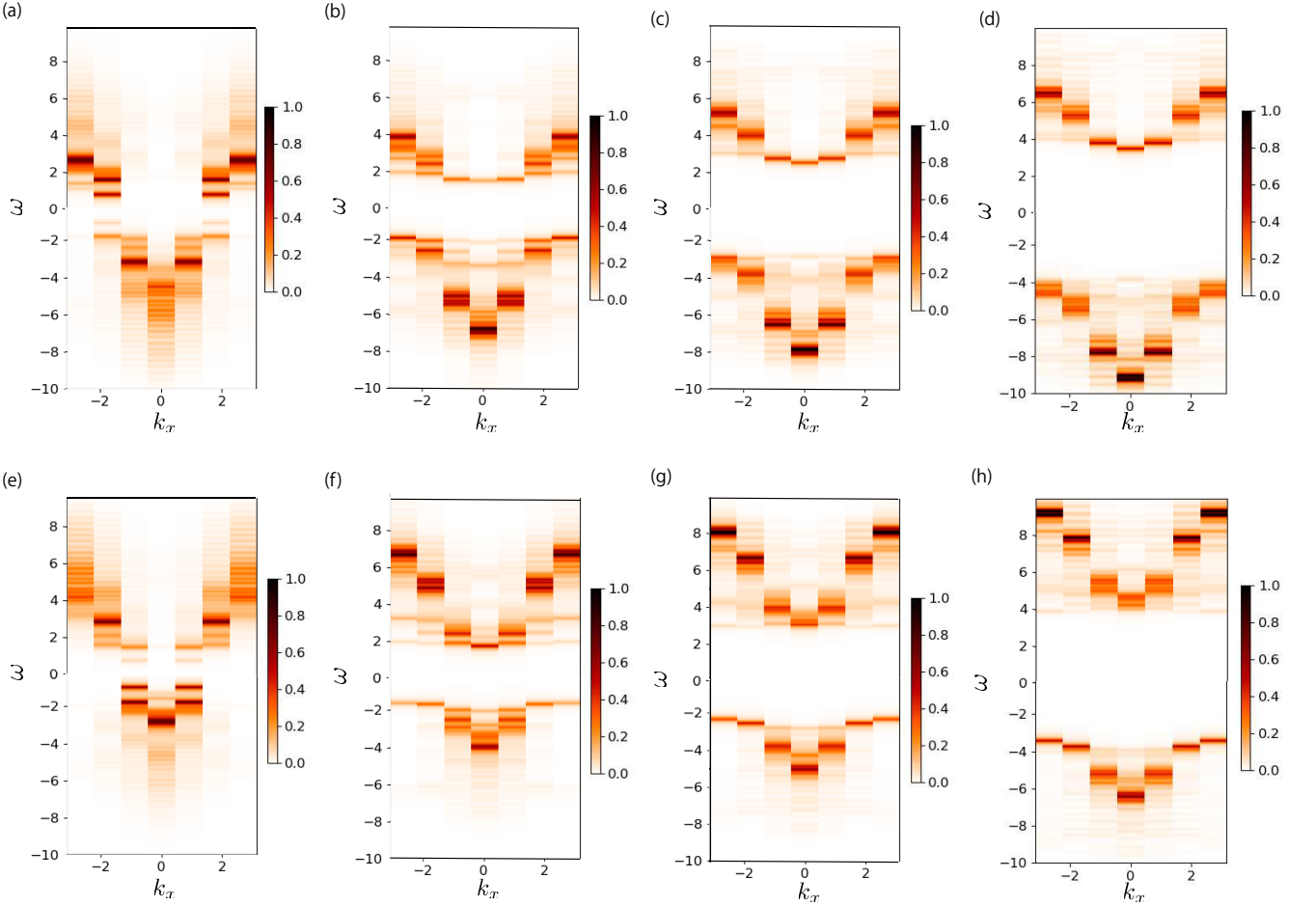}
    \caption{Same as Fig.~\ref{F4}, but the phase-string effect is removed. [(a) and (e)] $U=0$, [(b) and (f)] $U=4$, [(c) and (g)] $U=7$, and [(d) and (h)] $U=10$ calculated with $L_{x}=6$. Upper [(a)-(d)] and lower [(e)-(h)] figures are for $k_{y}=0$ and $\pi$ sectors, respectively.}
    \label{F5}
\end{figure*}

To further examine the effect of phase strings, we calculate the single-particle spectral function of the two-leg Hubbard ladder defined as
\begin{align}
A^{(e/h)}_{k_{y}}(\bm{k},\omega)=&-\frac{1}{\pi} \text{Im} G_{\uparrow}^{(e/h)}(k_{x},\omega)
\end{align}
with the Green function $G_{\uparrow}^{(e)}(\bm{k},\omega)$ and $G_{\uparrow}^{(h)}(\bm{k},\omega)$ for the electron-addition and electron-removal parts, respectively.
Each Green function is represented as
\begin{align}
G_{\sigma}^{(e)}(\bm{k},\omega)=&\langle \psi_{0} | c_{\bm{k},\uparrow} \cfrac{1}{\omega-(H-E_{0})+i\eta} c_{\bm{k},\uparrow}^{\dag} | \psi_{0} \rangle ,\\
G_{\sigma}^{(h)}(\bm{k},\omega)=&\langle \psi_{0} | c_{\bm{k},\uparrow}^{\dag} \cfrac{1}{\omega+(H-E_{0})+i\eta} c_{\bm{k},\uparrow} | \psi_{0} \rangle 
\end{align}
with
$
c_{\bm{k},\sigma} = \frac{1}{\sqrt{L}} \sum_{j_{x}=1}^{L_{x}} \sum_{j_{y}=0,1} e^{-ik_{x}j_{x}} e^{-ik_{y}j_{y}} c_{(j_{x},j_{y}),\sigma}.
$
Here, we take $\bm{k}=(k_{x},k_{y})$, where $k_{y}=0$ and $\pi$ corresponding to bonding and antibonding bands, respectively.
$c_{(j_{x},j_{y}),\sigma}$ is an electron annihilation operator of spin $\sigma$ at site $(j_{x},j_{y})$.
We define $A_{k_{y}}(k_{x},\omega)=A^{(e)}_{k_{y}}(k_{x},\omega) + A^{(h)}_{k_{y}}(k_{x},\omega)$ and $|\psi_{0}\rangle$ being the ground state of a Hamiltonian $H$ with energy $E_{0}$.
Using the Lanczos method, we calculate the Green functions via continued fraction expansion keeping the 100 Lanczos bases for $L_{x}=6$ with periodic boundary condition.

For $k_{y}=0$ sector, we show $A_{0}(k_{x},\omega)$ of the Hubbard model at half filling with $U=0$ in Fig.~\ref{F4}(a), $U=4$ in Fig.~\ref{F4}(b), $U=7$ in Fig.~\ref{F4}(c), and $U=10$ in Fig.~\ref{F4}(d).
For $k_{y}=\pi$ sector, we give $A_{\pi}(k_{x},\omega)$ with $U=0$ in Fig.~\ref{F4}(e), $U=4$ in Fig.~\ref{F4}(f), $U=7$ in Fig.~\ref{F4}(g), and $U=10$ in Fig.~\ref{F4}(h).
The color density depicts spectral weights as a function of momentum $k_{x}$ and frequency $\omega$.

Removing phase strings, we obtain $A_{0}(k_{x},\omega)$ in Figs.~\ref{F5}(a)-(d) and $A_{\pi}(k_{x},\omega)$ in Figs.~\ref{F5}(e)-(h).
We find that band dispersions substantially change by removing phase strings.
Even for $U=0$, $A_{0}(k_{x},\omega)$ of $\mathcal{H}_\text{r}$ [Fig.~\ref{F5}(a)] is not as simple as $A_{0}(k_{x},\omega)$ of $\mathcal{H}$ [Fig.~\ref{F4}(a)] with a Dirac-delta peak following cosine-like dispersion.
This is because $\mathcal{H}_\text{r}$ has a gauge interaction even for $U=0$.
We find that $A_{0}(k_{x},\omega)$ in Fig.~\ref{F5}(a) and $A_{\pi}(k_{x},\omega)$  in Figs.~\ref{F5}(b) of $\mathcal{H}_\text{r}$ has cosine-like dispersion with broad continuum.
Since $\mathcal{H}_\text{r}$ is not integrable even for $U=0$ in one-dimensional chain, the origin of broad continuum is not easily understood.
It is interesting to examine the integrable case, that is, the Schulz-Shastry model $\mathcal{H}_\text{SS}$.
The single-particle spectral function of $\mathcal{H}_\text{SS}$, which is obtained analytically~\cite{Penc2002}, has cosine-like dispersion with an additional broad continuum (see Appendix~\ref{Ac}).
We consider that this structure has the same origin with that found in $\mathcal{H}_\text{r}$.

\begin{figure}[t]
  \centering
    \includegraphics[clip, width=20pc]{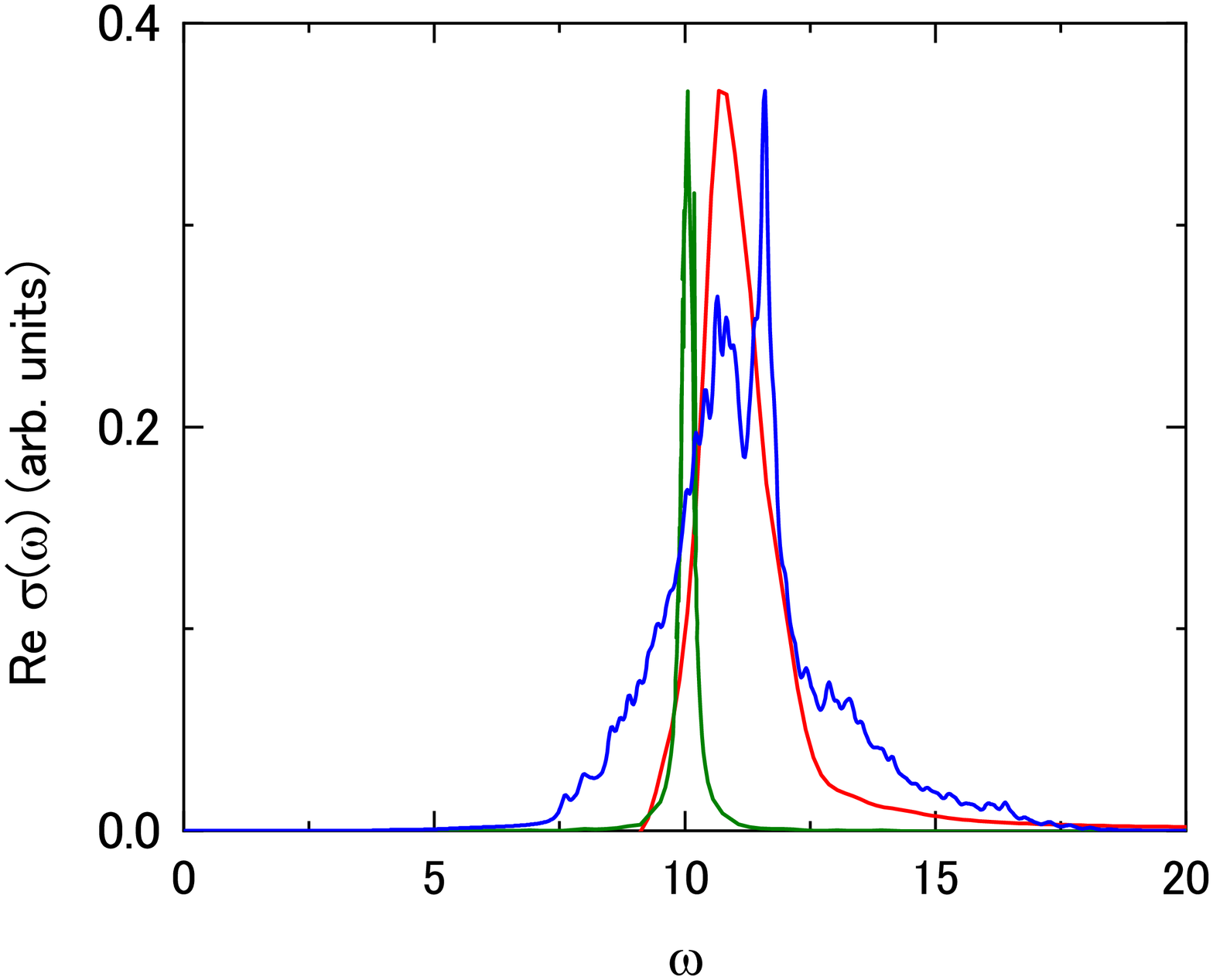}
    \caption{Comparison of $\text{Re}\sigma (\omega)$ at half-filling for $U=10$. The red line indicates $\text{Re}\sigma (\omega)$ of $\mathcal{H}_\text{r}$ calculated by the time-dependent DMRG with $L_{x}=20$. The blue line indicates $\text{Re}\sigma (\omega)$ of $\mathcal{H}_\text{r}$ obtained by Eq.~(\ref{Eq_ReSigma_approx}). The green line indicates $\text{Re}\sigma (\omega)$ of $\mathcal{H}$ in the one-dimensional chain with the Hubbard-I approximation. Note that $\text{Re}\sigma (\omega)$ indicated by the blue and green lines are normalized to have the same maximum values as that by the red line.}
    \label{F6}
\end{figure}

For large $U$, e.g., $U=10$, we find that the upper and lower Hubbard bands of $\mathcal{H}_\text{r}$ have cosine-like energy dispersion as shown in  Figs.~\ref{F5}(d) and \ref{F5}(h).
This is because string excitations associated with the motion of a hole is suppressed in $\mathcal{H}_\text{r}$ (see Appendix~\ref{Ad}).
The removal of phase strings eliminates a nontrivial U(1) phase acquired when a hole moves in AFM spin background, which suppresses string excitations emerging in single-hole dynamics.
As a result, a spin-polaron picture becomes valid and a quasiparticle with a cosine-like energy dispersion is composed.
If we completely ignore spin correlation, i.e., $\langle \bm{S}_{i}\cdot \bm{S}_{j}\rangle=0$, which is implicitly assumed~\cite{Dorneich2000, Grober2000} in the Hubbard-I approximation~\cite{Hubbard1968}, we obtain the Green function with a single pole following cosine-like dispersion in the upper and lower Hubbard bands.
This indicates that $S^{z}$ strings, which are all strings present in $\mathcal{H}_\text{r}$, do not qualitatively affect the energy dispersion of single-particle spectral functions.

Based on the results of calculated single-particle spectral functions, we qualitatively understand $\text{Re}\sigma(\omega)$ of $\mathcal{H}_\text{r}$ above the Mott gap.
We show $\text{Re}\sigma(\omega)$ of $\mathcal{H}_\text{r}$ at half filling by the red line in Fig.~\ref{F6}, which is obtained by the time-dependent DMRG for $U=10$ with $L_{x}=20$.
Compared with one-hole doped case shown in Fig.~\ref{F2}(e), we find a similar spectral structure above the Mott gap.
For simplicity, considering one-dimensional chain, where a current operator is represented as $j^{c}=\sum_{k_{x},\sigma} v_{k_{x}} c_{k_{x},\sigma}^{\dag}c_{k_{x},\sigma}$ with $v_{k_{x}}=-2t_\text{h}\sin(k_{x})$, we approximately obtain $\text{Re} \sigma (\omega)$ as~\cite{Pruschke1993}
\begin{align}\label{Eq_ReSigma_approx}
\text{Re} \sigma (\omega)\simeq&
-\frac{\pi}{L\omega} \sum_{k_{x}}  |v_{k_{x}}|^{2} \int_{-\infty} ^{\infty} d\omega' A_{0}(k_{x},\omega') \nonumber \\
&\times A_{0}(k_{x},\omega+\omega')  [f(\omega')-f(\omega+\omega')],
\end{align}
where $f(\omega)$ is the Fermi distribution function.
With $A_{0}(k_{x},\omega)$ shown in Fig.~\ref{F5}(d), we approximately obtain $\text{Re} \sigma (\omega)$ of $\mathcal{H}_\text{r}$ at half filling, which is shown in the blue line of Fig.~\ref{F6}.
Comparing the red and blue lines in Fig.~\ref{F6}, we find that $\text{Re}\sigma(\omega)$ obtained by Eq.~(\ref{Eq_ReSigma_approx}) captures narrow spectral feature in the red line of Fig.~\ref{F6}, although the peak position is slightly different from each other.

The green line of Fig.~\ref{F6} indicates $\text{Re} \sigma (\omega)$ of $\mathcal{H}$ for $U=10$ in a one-dimensional chain at half-filling obtained by the Hubbard-I approximation~\cite{Kubo1971}, where we completely ignore string excitations accompanied by the hopping of a hole.
If we take into account the string excitations, we obtain a broad continuum above the Mott gap.
The shape of $\text{Re} \sigma (\omega)$ becomes sharp if the contribution from string excitations is small.
Similarly, $\text{Re} \sigma (\omega)$ of $\mathcal{H}_\text{r}$ has a narrow peak as shown by the red line in Fig.~\ref{F6}.
This is caused by the reduction of string excitations, which is due to the removal of phase strings.
The contribution from $S^{z}$ strings remains, but they do not affect the formation of the narrow peak.

\section{Relation to the Floquet effective model}\label{sec_5}

Several ways have been suggested to control the phase-string effect~\cite{Zhu2016}.
For example, increasing spin polarization and introducing large hopping anisotropy~\cite{Liu2016} to reduce the phase-string effect have been suggested.
In this paper, we suggest another way to remove the phase-string effect via spin-sensitive periodic driving in the cold atom~\cite{Jotzu2015}.
The Hubbard model driven by spin-sensitive periodic electric field $\bm{\mathcal{A}^{\sigma}}(t)= \bm{\mathcal{A}_{0}^{\sigma}} \cos (\Omega t)$ with $\bm{\mathcal{A}_{0}^{\sigma}}=(\mathcal{A}_{0}^{\sigma},\mathcal{A}_{0}^{\sigma})$ is given by
\begin{align}
\mathcal{H}(t) =- \sum_{\langle i,j\rangle,\sigma} \left[t_\text{h} e^{-i\bm{\mathcal{A}^{\sigma}}(t)\cdot \bm{R}_{ij}} c_{i,\sigma} ^{\dag} c_{j,\sigma} + \text{H.c.}  \right] + \mathcal{H}^{I}.
\end{align}
Under near-resonant conditions $U=l\Omega \gg t_\text{h}$ with an integer $l$, photon-assisted spin-dependent correlated hopping~\cite{Keilmann2011,Greschner2014,Bermudez2015,Gorg2015} emerges in the Floquet effective Hamiltonian~\cite{Bukov2016,Shinjo2020} as
\begin{align}
\mathcal{H}_\text{eff}^{(0)}=\sum_{\langle i,j \rangle,\sigma}  
[
&-J_\text{eff}^{\sigma} (1-n_{i,-\sigma}) c_{i,\sigma}^{\dag} c_{j,\sigma} (1-n_{j,-\sigma}) \nonumber \\
&-J_\text{eff}^{\sigma} n_{i,-\sigma} c_{i,\sigma}^{\dag} c_{j,\sigma}  n_{j,-\sigma} \nonumber \\
&-K_\text{eff}^{\sigma} (-1)^{l} n_{i,-\sigma} c_{i,\sigma}^{\dag} c_{j,\sigma} (1-n_{j,-\sigma}) \nonumber \\
&-K_\text{eff}^{\sigma} (1-n_{i,-\sigma}) c_{i,\sigma}^{\dag} c_{j,\sigma} n_{j,-\sigma} \nonumber \\
&+ \text{H.c.}
],
\end{align}
where $J_\text{eff}^{\sigma}=t_\text{h}\mathcal{J}_{0}(\mathcal{A}_{0}^{\sigma})$ and $K_\text{eff}^{\sigma}=t_\text{h}\mathcal{J}_{l}(\mathcal{A}_{0}^{\sigma})$ with the $l$-th Bessel function of the first kind $\mathcal{J}_{l}$.
We may obtain a phase-string removed Hubbard model as a Floquet effective model if even $l$ is taken.
We consider $l=2$ for example.
Then, if we tune the parameters of external gauge field as $\mathcal{A}_{0}^{\uparrow}=0$, $\mathcal{A}_{0}^{\downarrow}=j_{1,m}$ with odd integer $m$, where  $j_{1,m}$ is $m$-th roots of $\mathcal{J}_{1}$, we obtain $\mathcal{H}_\text{eff}^{(0)}=\mathcal{H}_\text{r}^{T}$.
Even if we cannot strictly adjust a gauge field to this condition, we expect that it is still possible to reduce the effect of phase strings.

\section{summary and outlook}\label{sec_6}
We have studied the effect of string structure on the optical spectrum of hole-doped Mott insulators.
We have calculated the optical conductivity of the Hubbard and $t$-$J$ models by using time-dependent Lanczos and DMRG methods.
We have focused on the Mott insulators with a single hole and considered two-leg ladders, which are known to show MIR spectral weights.
Turning on and off the effect of phase strings, we have examined how these strings contribute to MIR weights.
We have found that MIR weights are crucially suppressed for both the Hubbard and $t$-$J$ models if we remove phase strings.
Although $S^{z}$ strings contribute to MIR weights, their contribution is smaller than that of phase strings.
This is because $S^{z}$ strings can be self-healed via quantum spin flips, while phase strings are not reparable.
Our findings indicate that a mutual Chern-Simons gauge field, which is an elementary force between spin and charge in the phase-string theory, is significant for generating MIR weights.
Conversely, if we remove this gauge field, a hole does not acquire a nontrivial U(1) phase when moving in AFM spin background, which gives rise to the significant reduction of string excitations emerging in single-hole dynamics.
As a result, a spin-polaron picture becomes valid and a quasiparticle with a cosine-like dispersion is recovered, which has been found by calculating single-particle spectra.
Furthermore, we have suggested a Floquet engineering to examine the phase-string effect in cold atoms.
Based on the phase-string theory, we have discovered a close relationship between single-hole dynamics and string excitations generated by involving spin and charge degrees of freedom.

In this paper, we have focused on two-leg ladder.
We consider that our findings are also valid for multi-leg ladders and two-dimensional clusters at least qualitatively since the effect of destructive interference due to phase strings is expected to exist in the presence of long-range order.
However, it is an open question as to whether the effect of $S^{z}$ strings on MIR weights is as small for two-dimensional systems as it is for two-leg ladders.
It is interesting to investigate the effects of phase strings on optical spectrum with various shapes of clusters, which remains as future work.

\begin{acknowledgments}
This work was supported by CREST (Grant No. JPMJCR1661), the Japan Science and Technology Agency, by the Japan Society for the Promotion of Science, KAKENHI (Grants No. 19H01829, No. JP19H05825, and No. 17K14148) from Ministry of Education, Culture, Sports, Science, and Technology (MEXT), Japan, and by MEXT HPCI Strategic Programs for Innovative Research (SPIRE; hp200071). Part of the numerical calculation was carried out using HOKUSAI at RIKEN Advanced Institute for Computational Science, the supercomputer system at the information initiative center, Hokkaido University, and the facilities of the Supercomputer Center at Institute for Solid State Physics, University of Tokyo.
\end{acknowledgments}

\appendix

\section{time-dependent Lanczos} \label{Aa}
To trace the temporal evolution of the system with small cluster, we employ the time-dependent Lanczos method to evaluate $|\psi (t)\rangle$~\cite{Park1986,Mohankumar2006,Giamarchi2016}.
Here, $|\psi(t+dt)\rangle\simeq\sum_{l=1}^{M}{e^{-i\epsilon_l dt}}|\phi_l\rangle\langle\phi_l|\psi(t)\rangle$, where $\epsilon_l$ and $|\phi_l\rangle$ are eigenvalues and eigenvectors of the tridiagonal matrix generated in the Lanczos iteration, respectively, $M$ is the dimension of the Lanczos basis, and $d t$ is the minimum time step. We set $M=50$ and $dt=0.02$.

\section{time-dependent DMRG} \label{Ab}
We briefly explain the time-dependent DMRG, which is used for obtaining the time evolution of the wave function of large cluster to which the Lanczos method cannot apply.
The dynamics of wave function $|\psi (t)\rangle$ of quantum systems is described by the time-dependent Schr\"{o}dinger equation, whose solution is given by
$|\psi (t)\rangle = U(t,0)|\psi(0) \rangle$,
where $|\psi (0)\rangle$ is the wave function at initial time $t=0$.
Here, 
\begin{align}
    U(t,0)=\hat T \exp \left[ -i\int _0 ^t ds H(s) \right]
\end{align}
is the time-evolution operator with the time-ordering operator $\hat T$ and the time-dependent Hamiltonian $H(t)$.
For small time step $dt$, in practice $dt=0.02$, we can approximate 
$U(t+dt,t)\simeq \exp [-idtH(t)]$.
To obtain $|\psi(t)\rangle$ accurately, we need to calculate $U(t+dt,t)$ as precise as possible.
One of the efficient approximations for $U(t+dt,t)$ is given by using the Suzuki-Trotter decomposition~\cite{White2004}.
However, this approach is basically restricted to one-dimensional case.
Another approach is the use of the kernel polynomial method to approximate $U(t+dt,t)$ as follows~\cite{Sota2007}:
\begin{align}
    U(t+dt,t) = \sum_{l=0}^{\infty} (-i)^l (2l+1)j_l(dt)P_l(H(t)) \nonumber \\
            \simeq \sum_{l=0}^{M_{p}} (-i)^l (2l+1)j_l(dt)P_l(H(t)),
\end{align}
where $j_l(s)$ is the spherical Bessel function of the first kind and $P_l(s)$ is the $l$-th Legendre polynomial.
They can be effectively obtained by the recurrence relations
\begin{align}
    j_{l+1}(x) = (2l+1)x^{-1}j_l(x) - j_{l-1}(x)
\end{align}
with $j_0(x)=x^{-1}\sin x$ and $j_1(x)=x^{-1}[-\cos x + x^{-1}\sin x]$ and
\begin{align}
    P_{l+1}(x) = \frac{2l+1}{l+1}xP_l(x) - \frac{l}{l+1}P_{l-1}(x)
\end{align}
with $P_0(x)=1$ and $P_1(x)=x$.
The calculation of the time-dependent DMRG in the present study is performed by using the kernel polynomial method with the truncation number $M_{p}$, practically for $M_{p}\approx 10$, which gives sufficiently converging result.
Furthermore, we use two target states $|\psi(t)\rangle$ and $|\psi(t+dt)\rangle$ in the time-dependent DMRG procedure to effectively construct a basis that can express wave functions in time-dependent Hilbert space.
With the two-target time-dependent DMRG procedure, we can calculate time-dependent physical quantities with high accuracy even when the Hamiltonian varies rapidly with time.

\section{Single-particle spectral function of the Schulz-Shastry model}\label{Ac}
\begin{figure}[t]
  \centering
    \includegraphics[clip, width=20pc]{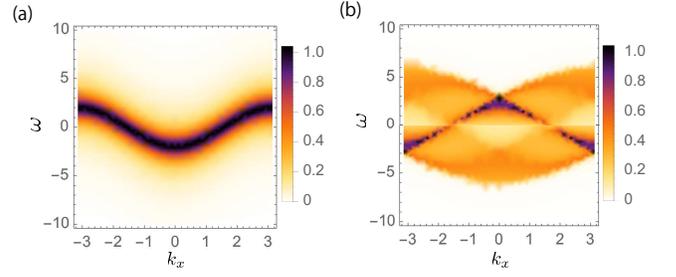}
    \caption{$ A(k_{x},\omega)$ of $\mathcal{H}_\text{SS}$ for $U=0$ in one-dimensional chain at half-filling. (a) $\phi=0$ and (b) $\phi=\pi$.}
    \label{F7}
\end{figure}

$\mathcal{H}_\text{SS}$ and $\mathcal{H}_\text{r}$ are similar to each other.
The only difference between the two models is the spin dependence of $\phi^{\sigma}$: $(\phi^{\uparrow},\phi^{\downarrow})=(\phi,-\phi)$ for $\mathcal{H}_\text{SS}$ and $(\phi^{\uparrow},\phi^{\downarrow})=(0,\phi)$ for $\mathcal{H}_\text{r}$.
Since $\mathcal{H}_\text{r}$ shares part of $\mathcal{H}_\text{SS}$, the spectral function of $\mathcal{H}_\text{r}$ may have a common feature with that of $\mathcal{H}_\text{SS}$.
In this section, we restrict ourselves to one-dimensional chain at half-filling.
Since only the $k_{y}=0$ sector is defined in one-dimensional chain, we take $ A(k_{x},\omega)=A_{0}(k_{x},\omega)$.
$ A(k_{x},\omega)$ of $\mathcal{H}_\text{SS}$, whose analytic form for $U=0$ in one-dimensional chain has been obtained~\cite{Penc2002}, has two possible components.
One is a Dirac-delta peak following the cosine-like dispersion, which is the remnant of the spectral function of the tight-binding model.
The other is a broader continuum.
We obtain only the former for $\phi=0$ as shown in Fig.~\ref{F7}(a) while only the latter for $\phi=\pi$ as shown in Fig.~\ref{F7}(b). 
In the intermediate $0< \phi <\pi$, we obtain the summation of them, that is, spectral weights following the cosine-like dispersion and additional broad continuum with momentum shift by $\phi$.
In the sense that there is cosine-like dispersion with an additional broad continuum, $ A(k_{x},\omega)$ of $\mathcal{H}_\text{r}$ has a common feature with that of $\mathcal{H}_\text{SS}$ with intermediate $\phi$ as far as $U=0$.

\section{The equation of motion for the Green function}\label{Ad}
Following Hubbard~\cite{Hubbard1968}, we split the electron annihilation operator into the two parts:
\begin{align}
c_{i,\sigma}=f_{i,\sigma} + g_{i,\sigma}
\end{align}
with
\begin{align}
f_{i,\sigma}=&\hat n_{i,-\sigma}c_{i,\sigma},\\
g_{i,\sigma}=& (1-\hat n_{i,-\sigma})c_{i,\sigma},
\end{align}
where $f$ and $g$ are eigenoperators of the interaction term of the Hubbard model since $[f_{i,\sigma},\mathcal{H}^{I}]=Uf_{i,\sigma}$ and $[g_{i,\sigma},\mathcal{H}^{I}]=0$ are realized.
Defining the Green function
\begin{align}
G_{\alpha,\beta}(\bm{k},\tau)=-\langle T \alpha_{\bm{k},\sigma}(\tau) \beta_{\bm{k},\sigma}^{\dag} \rangle
\end{align}
with $\alpha,\beta \in \{f,g\}$, the equation of motion is written as
\begin{align}
-\frac{\partial}{\partial \tau} G_{\alpha,\beta}(\bm{k},\tau)=\delta(\tau) \langle \{ \beta_{\bm{k},\sigma}^{\dag}, \alpha_{\bm{k},\sigma} \} \rangle - \langle T [ \alpha_{\bm{k},\sigma}(\tau),H] \beta_{\bm{k},\sigma}^{\dag} \rangle.
\end{align}
The commutators of $g$ and $\mathcal{H}^{T}$ are given as 
\begin{align}\label{eq_comm_up}
[g_{i,\uparrow},\mathcal{H}^{T}]=&-t_{h}\sum_{j\in \text{NN}(i)} \biggl[ c_{j,\uparrow} +(c_{j,\uparrow}S_{i}^{z}+c_{j,\downarrow}S_{i}^{-})\nonumber\\
&-\frac{1}{2} c_{j,\uparrow}\hat n_{i} + c_{j,\downarrow}^{\dag}c_{i,\downarrow}c_{i,\uparrow} \biggr]\nonumber \\
\simeq&-\frac{t_{h}}{2}\sum_{j\in \text{NN}(i)}c_{j,\uparrow}
+\xi_{i,\uparrow}
-t_{h}\sum_{j\in \text{NN}(i)}c_{j,\downarrow}^{\dag}c_{i,\downarrow}c_{i,\uparrow},
\end{align}
where 
\begin{align}
\xi_{i,\uparrow}=&\sum_{j\in \text{NN}(i)} ( c_{j,\uparrow}S_{i}^{z}+c_{j,\downarrow}S_{i}^{-})
\end{align}
is a spin-$\frac{1}{2}$ string operator.
In the last line of Eq.~(\ref{eq_comm_up}), we assume that the system is at nearly-half filling, i.e., $\langle \hat n_{i,\uparrow}\rangle =\langle \hat n_{i,\downarrow}\rangle \simeq\frac{1}{2}$ and ignore charge fluctuations, leading to the replacement of $\hat n_{i}$ with $1$.
The second term of Eq.~(\ref{eq_comm_up}) represented by $\xi_{i,\uparrow}$ not only creates a hole on site $j$ but dresses this hole with a spin excitation on a neighboring site, which is closely related to string excitations accompanied by the hopping of a hole.
The third term describes the coupling of the hole to the pairing excitation.
For a large $U>0$ near half-filling, $\xi_{i,\uparrow}$ contributes significantly, while the pairing-excitation term does not.

Similarly, we obtain commutators for the phase-string-removed Hamiltonian $\mathcal{H}_{\text{r}}^{T}$ as
\begin{align} \label{eq_comm_up_r}
[g_{i,\uparrow},\mathcal{H}_{\text{r}}^{T}]
=&-t_{h}\sum_{j\in \text{NN}(i)} \Biggl\{ \biggl[ c_{j,\uparrow} - c_{j,\uparrow}\left(\frac{\hat n_{i}}{2}-S_{i}^{z}\right)  \biggr]\nonumber \\
&+(1-2\hat n_{j,\uparrow})  \biggl[ c_{j,\downarrow}S_{i}^{-}-c_{j,\downarrow}^{\dag}c_{i,\downarrow}c_{i,\uparrow}\biggr] \Biggr\} \nonumber \\
\simeq& -\frac{t_{h}}{2}\sum_{j\in \text{NN}(i)}   c_{j,\uparrow}  + \tilde \xi_{i,\uparrow},
\end{align}
where
\begin{align}
\tilde \xi_{i,\uparrow}=&\sum_{j\in \text{NN}(i)} c_{j,\uparrow}S_{i}^{z}.
\end{align}
In the final line of Eq.~(\ref{eq_comm_up_r}), we assume that the system is at nearly-half filling, i.e., $\langle \hat n_{i,\uparrow}\rangle =\langle \hat n_{i,\downarrow}\rangle \simeq\frac{1}{2}$ and ignore charge fluctuations, leading to the replacement of $\hat n_{i}$ and $1-2\hat n_{i,\uparrow}$ with $1$ and $0$, respectively.
If we compare spin-$\frac{1}{2}$ string operators $\xi_{i,\uparrow}$ with $\tilde \xi_{i,\uparrow}$, we find that scattering process related to transverse component of spin $S_{i}^{-}$ disappears for phase-string-removed model.

The structure of energy dispersion significantly depends on whether spin excitations accompanied by the hopping of a hole are incorporated into $\xi$ or $\tilde \xi$.
For $U=10$, we obtain the cosine-like band dispersion as found in Figs.~\ref{F5}(d) and \ref{F5}(h) for $\mathcal{H}_\text{r}$, which corresponds to the incorporation of spin excitations by $\tilde \xi$, while we do not in Figs.~\ref{F4}(d) and \ref{F4}(h) for $\mathcal{H}$ generating $\xi$.
If no spin excitation is incorporated, i.e., spin-$\frac{1}{2}$ string operators are taken as 0, we obtain the Green function as
\begin{align}\label{eq_Green_HubbardI}
G(k_{x},\omega)=\frac{1}{\omega-\varepsilon_{k_{x}}-\Sigma(\omega)}
\end{align}
with the self energy 
\begin{align}
\Sigma(\omega)=\frac{\langle \hat n_{i} \rangle}{2}U+\frac{\langle \hat n_{i} \rangle}{2}\left(1-\frac{\langle \hat n_{i} \rangle}{2}\right)\frac{U^{2}}{\omega-\left(1-\frac{\langle \hat n_{i} \rangle}{2}\right)U}.
\end{align}
This is the Green function obtained by the Hubbard-I approximation, where $\langle \bm{S}_{i}\cdot \bm{S}_{j}\rangle=0$ is implicitly assumed~\cite{Dorneich2000, Grober2000}.
This Green function has a single pole following cosine-like dispersion in the upper and lower Hubbard bands.
We therefore find a cosine-like dispersion in the single-particle spectral functions for both cases where the spin-$\frac{1}{2}$ string operators are 0 and $\tilde \xi$.
This indicates that $S^{z}$ strings, which are all strings present in $\mathcal{H}_\text{r}$, do not substantially affect the structure of energy dispersion of single-particle spectral functions.
In contrast, the effect of phase strings is so large that the dispersions change qualitatively.

\end{document}